# Extrapolation of neutron-rich isotope cross-sections from projectile fragmentation

M. Mocko,[1] M. B. Tsang,[1][•] Z.Y. Sun,[1,2] L. Andronenko,[1] M. Andronenko,[1] F. Delaunay,[1] M. Famiano,[1] W. A. Friedman,[3] V. Henzl,[1] D. Henzlova,[1] H. Hui,[1] X. D. Liu,[1] S. Lukyanov,[1] W. G. Lynch,[1] A. M. Rogers,[1] and M. S. Wallace[1]

[1] *National Superconducting Cyclotron Laboratory and Department of Physics & Astronomy, Michigan State University, East Lansing, Michigan 48824, U*SA*
[2] *Institute of Modern Physics, CAS, Lanzhou 730000, China*
[3] *Department of Physics, University of Wisconsin, Madison, Wisconsin 53706, USA*

## Abstract

Using the measured fragmentation cross sections produced from the $^{48}$Ca and $^{64}$Ni beams at 140 MeV per nucleon on $^{9}$Be and $^{181}$Ta targets, we find that the cross sections of unmeasured neutron rich nuclei can be extrapolated using a systematic trend involving the average binding energy. The extrapolated cross-sections will be very useful in planning experiments with neutron rich isotopes produced from projectile fragmentation. The proposed method is general and could be applied to other fragmentation systems including those used in other radioactive ion beam facilities.

---

- Corresponding author: Tsang@nscl.msu.edu

# I. Introduction

With the availability of radioactive ion beams, the frontier of nuclear science moves towards the understanding of the properties of rare isotopes [1]. Such studies often require effective means to produce rare isotopes using projectile fragmentation. In this paper we will discuss a parameterization of the exponential dependence of the projectile fragmentation cross section on the average binding energy, which allows extrapolations to predict the yields of the most neutron-rich isotopes.

Calculations of production rates at current radioactive ion beam facilities such as the National Superconducting Cyclotron Laboratory at Michigan State University and next generation radioactive ion beam facilities mainly rely on the use of an empirical code, EPAX [2]. The current parameters in EPAX were obtained by fitting fragmentation cross sections from reactions at incident energy greater than 200 MeV per nucleon [2]. At that energy, limiting fragmentation, which assumes that the reaction mechanism does not depend on incident energy or target, is a reasonable assumption. However, the validity of this assumption diminishes with decreasing incident energy. In addition, EPAX predictions become increasingly inaccurate as one diverges from the valley of beta-stability [3, 4]. Optimizing the EPAX formulae by fitting one specific reaction improves the accuracies of predicted cross-sections for most nuclei but the uncertainties associated with extrapolation to very neutron rich nuclei remain large [5].

Extrapolation to unmeasured cross-sections is most accurate if the dependence is linear or exponential. We found such exponential dependence in the average binding energy. To improve the accuracy of the neutron rich isotopes, we restrict our fits to the isotopes in a single element produced in a single reaction. The fits are then used to extrapolate the abundances of neutron rich isotopes with very low cross-sections that may not have been observed. In this respect, it differs from the global nature of EPAX.

## II. Cross-section systematics

In a recent experiment, the cross-sections of 202 and 243 isotopes have been measured in the projectile fragmentation of $^{48}$Ca+$^9$Be and $^{64}$Ni+$^9$Be systems respectively [3,4]. Similar numbers of fragments have been measured for the $^{48}$Ca+$^{181}$Ta and $^{64}$Ni+$^{181}$Ta reactions [3,4]. These accurate measurements provide a unique data set for testing different approaches to predicting the production rates of rare isotopes.

It is well known that the average binding energy obtained by dividing the binding energy by the number of nucleons as a function of mass number yields a broad peak around A~60 region [6]. A similar peak is observed in the average binding energy of isotopes of the same element. As an example, the bottom panel of Figure 1 shows the average binding energy [7] as a function of mass number A for the silicon isotopes produced in the fragmentation of $^{48}$Ca+$^9$Be reactions [3,4]. The odd and even effects of the binding energy can be clearly seen and the peak value of <B>~8.5 MeV occurs around $^{30}$Si. Over the range of Si isotopes from A= 26-41, the average binding energy only changes by about 1 MeV and a linear scale for <B> is used.

Interestingly, the general trend of the mass dependence of the average binding energy is mirrored in the isotope cross-section distributions [3]. A corresponding plot for Si isotopes is shown in the top panel of Figure 1. Unlike the small change in <B>, however, the cross-sections change by more than seven order of magnitudes and to display this trend, they must be plotted on a log scale. The similarity in the two plots suggests an exponential dependence of the cross-sections on the average binding energy <B>=B/A,

$$\sigma = C exp(<B>/\tau_i). \qquad (1)$$

where C is the normalization constant and $\tau_i$ is the "inverse slope". To avoid confusion regarding the convention used for the sign of the binding energy, we define B to be positive.

In this paper, we focus mainly on the empirical relationship observed. At present, there is no simple theoretical justification for Eq. (1). Models that calculate the excitation of the projectile in the collision and follow its subsequent decay predict that the final yields reflect a complex interplay of reaction dynamics and statistical decay [8,9]. Near the drip-lines, the yields require the description of very low probability ($<10^{-8}$) fluctuations in the production mechanism, which is a goal that is presently outside the reach of such models. Instead, we take Eq. 1 as an empirical starting point and explore how it can be used and search for its limitations.

To test the validity of Eq. (1), we plot the aluminum isotope cross-sections as a function of the average binding energy <B> in Figure 2. We choose odd Z isotopes to minimize the odd-even effects. The mass numbers of the isotopes are labeled next to the symbols. $^{34,35,36,37}$Al isotopes seem to follow Equation (1). (The uncertainty in the $^{38}$Al cross section ($\approx$40%) stems from an incomplete measurement of the momentum distribution; it is possible that the extrapolation to unmeasured momenta may be underestimated [4].) Lighter masses of A$\approx$26-30 (closed diamonds), corresponding to the peak of the isotopic distribution deviate significantly from the trend of the heavier masses. If one follows the trend of $^{34,35,36,37}$Al to lighter masses of about A$\approx$31, one can discern a slight curvature to the data.

To reduce the sensitivity to the curvature near the peak of the isotopic yields, we have extrapolated the yields for the $^{48}$Ca systems by fitting the heaviest neutron-rich isotopes with <B> less than 8 MeV except for $^{38}$P and $^{39}$P nuclei. Figure 3 shows neutron-rich isotope cross-sections plotted as a function of the average binding energy <B> after correcting for pairing energy (see below). The cross-sections of the odd Z isotopes (Z=11, 13, 15) and even Z neutron isotopes (Z=10, 12, 14) are plotted on the top and bottom panels, respectively. The heaviest isotope plotted in each element is labeled in the figure. The exponential dependence of the systematics seems to be fairly accurate over a moderately wide range of neutron-rich nuclei, providing an empirical means to extrapolate the cross sections of even more neutron-rich nuclei.

In the case of fragmentation of $^{64}$Ni beams, most of the neutron-rich fragments produced with mass near 60 are in the region of nuclei with large average binding energy. The change of <B> over an isotope chain is much smaller than those observed in fragments produced in the fragmentation of $^{48}$Ca beams. The odd-even effects are larger as shown in the left panel of Figure 4 where the measured cross-sections of $^{55-60}$Cr isotopes are plotted as a function of B/A. Such effects, arising from the influence of pairing on the energies and level densities of these nuclei, are readily observed in even-even nuclei.

### III. Extrapolation of cross-sections based on systematics

In order to extrapolate to unmeasured cross-sections of the more neutron rich isotopes, we fit the measured cross-sections of the most neutron-rich isotopes of each element with a modified form of Eq. (1).

$$\sigma = C exp((<B> - \varepsilon_{pair} - 8)/\tau). \qquad (2)$$

where $\varepsilon_{pair} = \kappa \varepsilon \cdot A^{-7/4}$ is the pairing correction factor [6] in MeV and $\kappa$=0, 1 and -1 for odd-even, even-even and odd-odd isotope, respectively. The pairing correction is not strictly given by the ground state pairing energy. It also reflects the number of particle stable states in the observed fragments, an effect that can counteract the effect of pairing on the ground state energy. The reduction of the odd-even effect we obtained by including a pairing correction, **$\varepsilon_{pair}$**, can be seen in the right panel of Figure 4. To avoid excessively large values arising from the exponential term, a constant of 8 MeV is subtracted from <B> in Eq. (2). The lines shown in Figure 3 and 4 are the best fits over the range of isotopes plotted using Eq. (2).

Table 1 lists the fitting parameters of Z=10-17 for $^{48}$Ca+$^{9}$Be and $^{48}$Ca+$^{181}$Ta reactions. Quality of the fits for both reactions is similar. $\varepsilon$ are larger for even Z nuclei and generally increase with Z. We set **$\varepsilon \geq 0$** to be the same for the Be and Ta targets. The values for $\tau$ are similar for both targets and decrease with increasing Z. The absolute value of C depends on the value of (<B>-8) and varies with elements. C mainly reflects the relative isotope yields between the two targets with the fragment yields (and C values) being higher in Ta target especially for lighter fragments [3,4].

Table 2 contains the fitting parameters of isotopes with Z=20-25 for $^{64}$Ni+$^{9}$Be and $^{64}$Ni+$^{181}$Ta reactions. The fits for lighter fragments are not included in Table 2 because most neutron-rich isotopes with Z<20 have higher cross-sections from the fragmentation of $^{48}$Ca and would most likely be measured with the latter reaction. The properties of the fitting parameters are similar to those from the fragmentation of $^{48}$Ca except that the values for τ are nearly constant within the narrow range of Z we study.

With the fitting parameters listed in Table 1 and 2, cross-sections of unmeasured neutron-rich isotopes can be obtained. Furthermore, the systematics can also be used to calculate the cross-sections of undiscovered nuclei using the estimated binding energy. One nucleus of particular interest is $^{40}$Mg as its non-existence will determine the neutron drip line for N=28 isotones. At present only the neutron drip line for Z≤8 has been determined definitively. From our extrapolation, the predicted cross-sections for $^{40}$Mg are (1-2)x10$^{-11}$ mb for the $^{48}$Ca+$^{9}$Be reaction and (4-8)x10$^{-11}$ mb for the $^{48}$Ca+$^{181}$Ta reactions using the binding energy range listed in ref [7]. It should be noted that prediction of the cross-sections of $^{40}$Mg nucleus requires extrapolation of the current systematics to six additional neutrons over six order of magnitude of the cross-sections. Most of the fits in the present work are performed with 4-5 nuclei. It is therefore important that the proposed algorithm be verified experimentally with a longer chain of neutron rich isotopes of one element and a model be developed to understand the observed exponential behavior. The current systematics with the fitting parameters together with the measured cross section data [3, 4] is adequate to estimate the rates of many neutron rich isotopes used as secondary beams at the National Superconducting Cyclotron Laboratory.

## VI. Summary

We have shown that the cross-sections of neutron-rich isotopes produced from the projectile fragmentation of $^{48}$Ca and $^{64}$Ni beams with two commonly used targets of $^{9}$Be and $^{181}$Ta can be predicted using a simple systematics involving the average binding energy of the isotopes for each element. To facilitate the production of secondary beams, the observed systematics of the neutron-rich isotope cross-sections suggests that comprehensive fragmentation cross-sections of the most commonly used primary beams should be measured in each facility. The investment is modest, usually 1-2 days of beam time but the benefits of planning experiments with accurate beam rates are invaluable.

**ACKNOWLEDGEMENTS**

This work is supported by the National Science Foundation under Grant Nos. PHY-01-10253, PHY-0606007.

**REFERENCES:**

[1] D.F. Geesaman, C.K. Gelbke, R.V.F. Janssens, and B.M. Sherrill Annu. Rev. Nucl. Part. Sci. 56 (2006) 53 and references therein.
[2] K. Summerer and B. Blank, Phys. Rev. C **61** (2000) 034607.
[3] M. Mocko et al., Phys. Rev. C, **74** (2006) 054612.
[4] M. Mocko, "Rare Isotope Production", PhD thesis, Michigan State University (2006).
[5] M. Mocko et al., NSCL preprint, MSUCL1347 (2007).
[6] R.R. Roy and B.P. Nigam, Nuclear Physics, John Wiley & Sons, Inc. (1967), Pg. 14-152.
[7] G.Audi, A.H.Wapstra and C.Thibault, Nucl. Phys. A729 (2003) 337.

http://csnwww.in2p3.fr/amdc/masstables/Ame2003/mass.mas03. All the binding energy values and their associated uncertainties used in this work have been obtained from this reference.
[8] J.-J. Gaimard and K.-H. Schmidt, Nucl. Phys. A **531**, 709-745 (1991).
[9] W. A. Friedman and M.B. Tsang, Phys. Rev. C67 (2003) 051601.

Table 1: Fitting parameters for Eq. (2) obtained from fitting neutron-rich nuclei in Z=10-17 for both $^{48}$Ca+$^{181}$Ta and $^{48}$Ca+$^9$Be reactions.

| Z | $^{48}$Ca + $^{181}$Ta | | | $^{48}$Ca + $^9$Be | | |
|---|---|---|---|---|---|---|
| | C (mb) | ε (MeV) | τ (MeV) | C (mb) | ε (MeV) | τ (MeV) |
| 10 | 2.57 | 0 | 0.0786 | 0.805 | 0 | 0.0803 |
| 11 | 1.57 | 6 | 0.0838 | 0.644 | 6 | 0.0812 |
| 12 | 0.423 | 6 | 0.0629 | 0.180 | 6 | 0.0616 |
| 13 | 0.272 | 0 | 0.0643 | 0.117 | 0 | 0.0635 |
| 14 | $3.05\times10^{-2}$ | 5 | 0.0600 | $1.40\times10^{-2}$ | 5 | 0.0549 |
| 15 | $1.43\times10^{-2}$ | 10 | 0.0561 | $6.31\times10^{-3}$ | 10 | 0.0545 |
| 16 | $2.02\times10^{-3}$ | 16 | 0.0496 | $8.57\times10^{-4}$ | 16 | 0.0482 |
| 17 | $1.29\times10^{-3}$ | 18 | 0.0480 | $9.41\times10^{-4}$ | 18 | 0.0500 |

Table 2: Fitting parameters for Eq. (2) obtained from fitting neutron-rich nuclei in Z=20-25 for both $^{64}$Ni+$^{181}$Ta and $^{64}$Ni+$^9$Be reactions.

| Z | $^{64}$Ni + $^{181}$Ta | | | $^{64}$Ni + $^9$Be | | |
|---|---|---|---|---|---|---|
| | C (mb) | ε (MeV) | τ (MeV) | C (mb) | ε (MeV) | τ (MeV) |
| 20 | $1.27\times10^{-11}$ | 13 | 0.0286 | $3.90\times10^{-12}$ | 13 | 0.0280 |
| 21 | $9.35\times10^{-11}$ | 7 | 0.0301 | $1.78\times10^{-11}$ | 7 | 0.0286 |
| 22 | $2.30\times10^{-10}$ | 16 | 0.0322 | $5.26\times10^{-11}$ | 16 | 0.0307 |
| 23 | $4.58\times10^{-10}$ | 9 | 0.0318 | $1.50\times10^{-10}$ | 9 | 0.0307 |
| 24 | $9.01\times10^{-11}$ | 25 | 0.0297 | $6.10\times10^{-11}$ | 25 | 0.0296 |
| 25 | $8.96\times10^{-11}$ | 12 | 0.0291 | $1.57\times10^{-10}$ | 12 | 0.0300 |

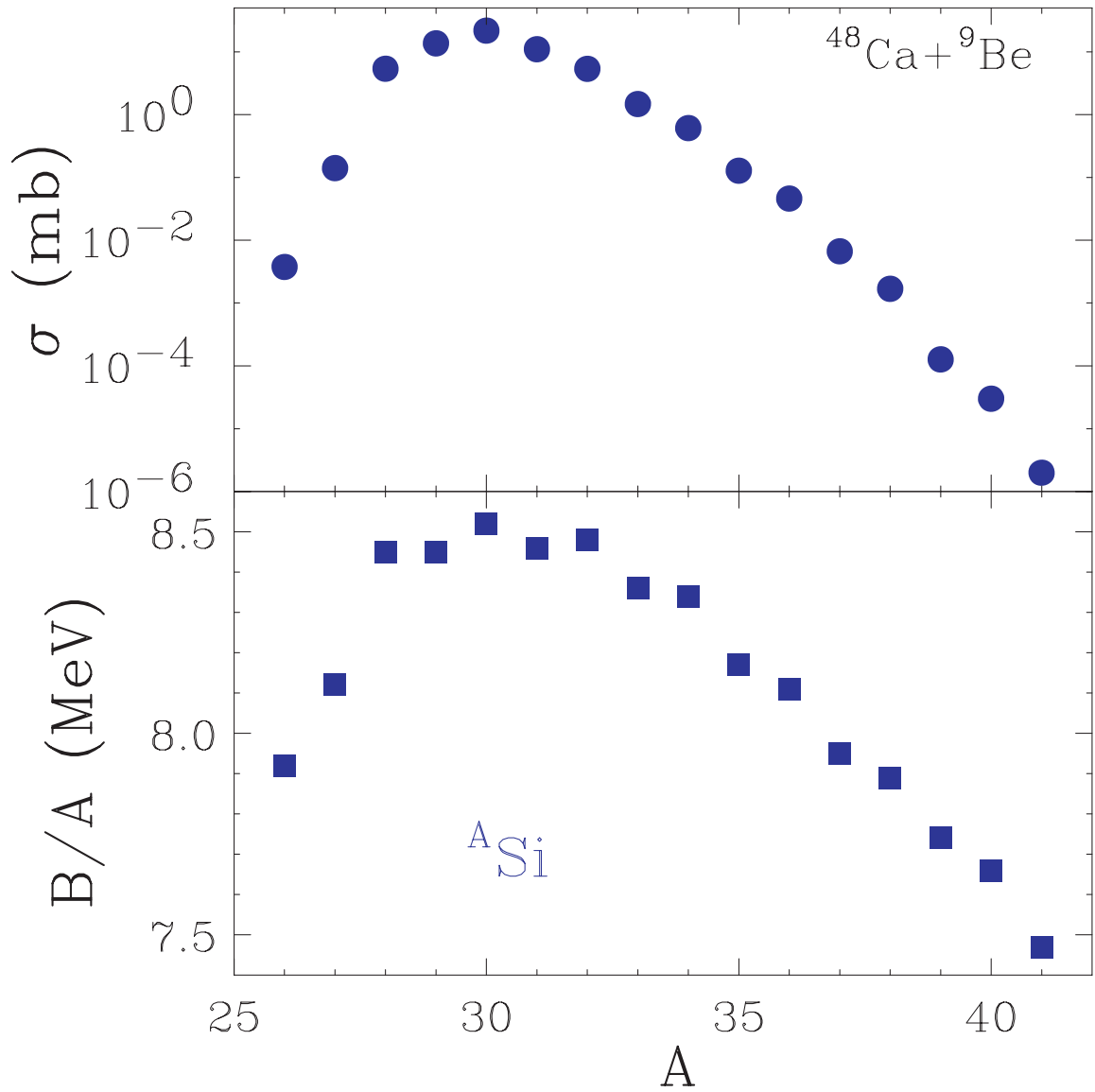

FIG 1: (Color online) Fragment cross section [3, 4] (top panel) and average binding energy [7] (bottom panel) plotted as a function of mass number for silicon isotopes.

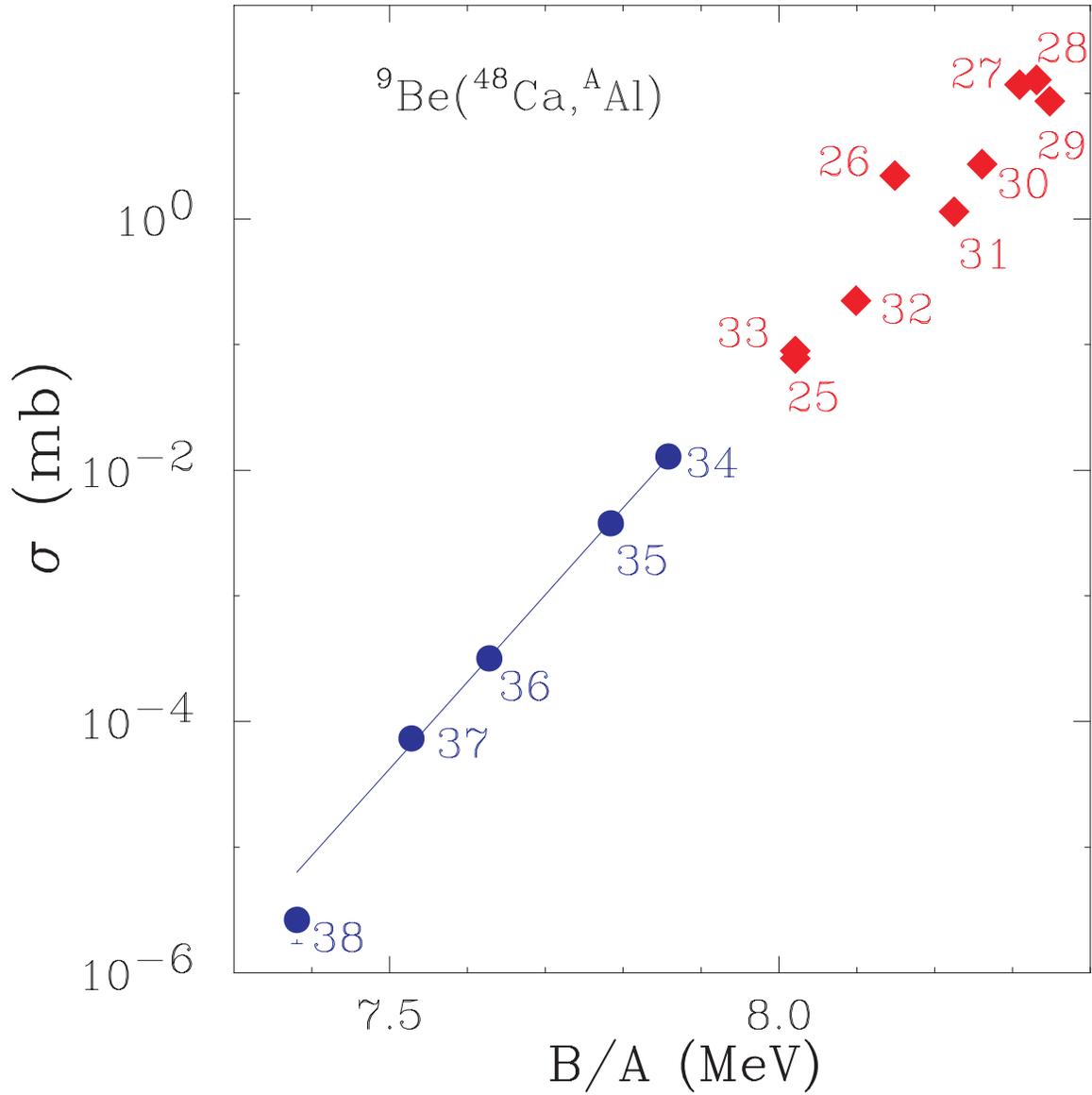

FIG 2: (Color online) Fragmentation cross sections of aluminum isotopes produced from $^{48}$Ca+$^9$Be reaction, plotted as a function of binding energy per nucleon. The mass numbers are labeled next to the symbols. The circles represent the five heaviest aluminum isotopes, $^{34}$Al to $^{38}$Al and the line is the best fits of $^{34-37}$Al isotopes using Eq. (2). The lighter aluminum isotopes are denoted by the closed diamonds.

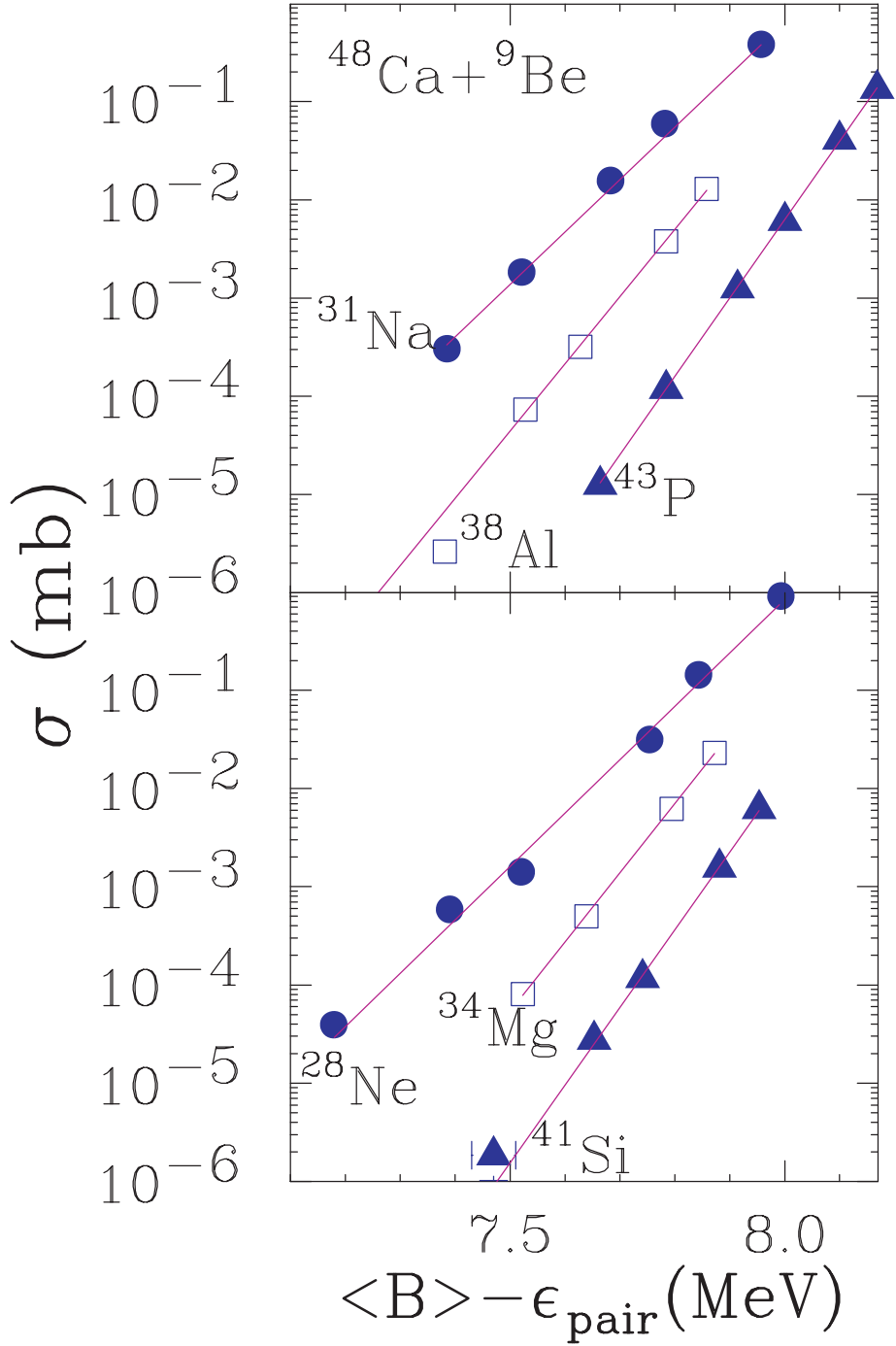

FIG 3: (Color online) Fragmentation cross sections of neutron rich isotopes of Na, Mg, Al, Si, P, and S elements produced from $^{48}$Ca+$^9$Be reaction, plotted as a function of binding energy per nucleon with pairing corrections. Odd Z isotopes are plotted in the top panel and even Z isotopes in the bottom panel. Lines show the best fits of Eq (2) using the parameters listed in Table 1.

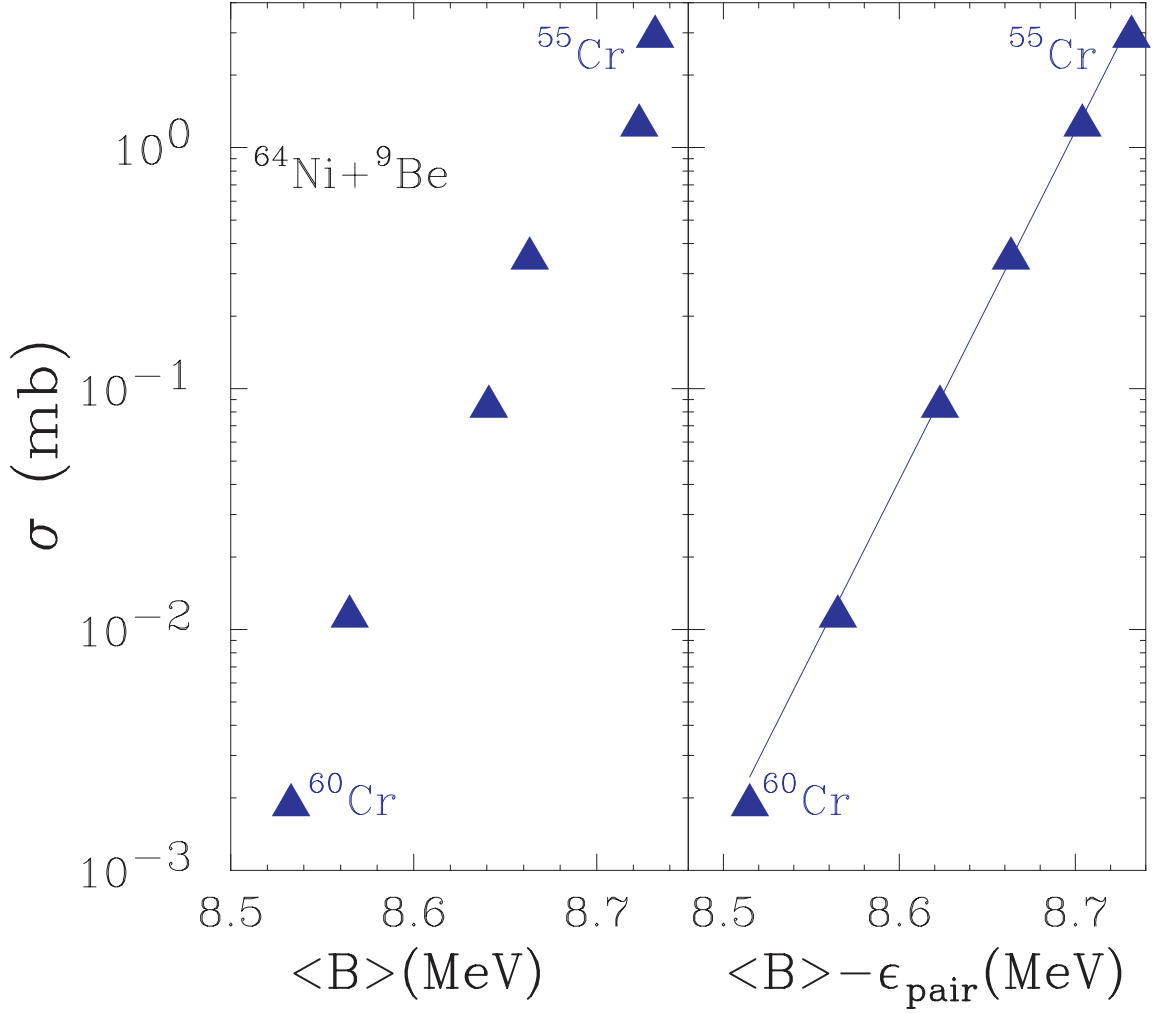

FIG 4: (Color online) Fragmentation cross section plotted as a function of average binding energy for neutron-rich chromium isotopes produced from the fragmentation of $^{64}$Ni+$^9$Be reaction. In the right panel, pairing energy has been subtracted from $^{56,58,60}$Cr isotopes to minimize the odd-even effects. The line is the best fit through the data points using Eq. (2) with the parameters listed in Table 2.